\documentclass[12pt]{article}

\begin{document}
\begin{center}

{\bf{\Large Family Of Rotating Anisotropic Fluid Solutions which Match to Kerr's Solution
}}

\vspace{1cm}
E. Kyriakopoulos\footnote{E-mail: kyriakop@central.ntua.gr}\\
Department of Physics\\
National Technical University\\
157 80 Zografou, Athens, GREECE
\end{center}

\begin {abstract}

We present a family of exact rotating anisotropic fluid solutions, which satisfy all energy conditions for certain values of their parameters. The components of the Ricci tensor $R_{\mu\nu}$ the eigenvalues of the tensor $R_\mu^\nu$ and the energy-momentum tensor $T_{\mu\nu}$ of the solutions are given explicitly. All members of the family have the ring singularity of Kerr's solution and most of them one or two more singularities. The solutions can be matched to the solution of Kerr on three closed surfaces, which for proper values of the parameters of the solutions approximate oblate spheroids. All matching surfaces are thin shells. For some values of a constant the surface density in one of them is positive everywhere and in this surface and in its interior all energy conditions are satisfied.
\\

PACS number(s): 04.20.-q,  04.20.Jb

\end {abstract}

Local isotropy is usually assumed in general relativity. However some phenomena seem to lead to local anisotropy. Since the remarkable work of Browers and Liang \cite{Bo1}, which refers to the non-rotating case , the influence of local anisotropy in general relativity has been extensively studied \cite{He1}. This study was extended to the case of rotation \cite{Gu1}.

In astrophysics it is very important to have a metric for the interior of a rotating star, which should match to the solution of Kerr \cite{Ke1}, since many times we take the solution of Kerr as exterior solution. Many people tried to find interior solutions but what they found had in most cases some problem. For example the matching was approximate \cite{Co1}-\cite{Fl2}. Also a family of rotating disks was given, the energy-momentum tensor of the family
was analyzed and it was found that none of the disks satisfy the dominant energy conditions \cite{Ha2}. However relativistic disks as sources of the Kerr metric with physical energy-momentum tensors have been found \cite{Bi1}. The question if a perfect fluid can be the source of the metric of Kerr is an open one \cite{Kr1},\cite{Mc1}.

In this paper we present a family of rotating anisotropic fluid solutions, which depends on five parameters including the parameters of mass and angular momentum and which for certain values of these parameters satisfy the dominant the strong and of course the weak energy conditions. The non vanishing components of the Ricci tensor $R_{\mu\nu}$ and the eigenvalues of $R_\mu^\nu$ are given explicitly. All solutions are singular on a ring lying in the equatorial plane [the ring singularity of Kerr's solution], while most of them have one or two more singularities  The infinite red shift surfaces of the solutions are given.

The solutions of the family can be matched to the solution of Kerr on three surfaces, which for proper values of the parameters of the solutions approximate oblate spheroids. Since not all components of the second fundamental form are zero the surfaces are thin shells. For two of them the surface energy tensor $S_{ij}$ is computed and then the surface density $\sigma(\theta)$ is calculated. It is found that for some values of a constant the surface density in one of these two surfaces is positive everywhere and also in this surface and in its interior all energy conditions are satisfied.

We consider a metric which in Boyer-Lindquist coordinates \cite{Bo2} has the form
\[g_{\mu\nu}=\{[-T, 0, 0, -a(1-x^2)(1-T)], [0, \frac{f(r,x)}{\rho^2T+a^2(1-x^2)}, 0, 0],\]
\begin{equation}
[0, 0, f(r,x), 0], [-a(1-x^2)(1-T), 0, 0, (1-x^2)(r^2+a^2x^2+a^2(1-x^2)(2-T))]\}
\label{1}
\end{equation}
where a is an arbitrary constant,
\begin{equation}
x=\cos\theta,\>\>\>\>\>\>\>\> \rho^2=r^2+a^2x^2, \>\>\>\>\>\>\>\>T=1+\frac{h(r)}{\rho^2}
\label{2}
\end{equation}
and h(r) and f(r,x) are functions to be determined. The above form of $g_{\mu\nu}$ approaches the general form of a metric we have considered before \cite{Ky1} and becomes Kerr's metric for $h(r)=-2Mr$ and $f(r,x)=\rho^2$. Starting from a metric of the above form we calculate \cite{Bo3} the Ricci tensor $R_{\mu\nu}$ and the expressions from which the eigenvalues of the matrix $R_\mu^\nu$ are obtained. To get an anisotropic fluid solution the eigenvalues $\lambda_\theta$ and $\lambda_\phi$ of the matrix $R_\mu^\nu$ must be equal. Imposing the condition $\lambda_\theta=\lambda_\phi$ we calculate \cite{Ky2} the functions $h(r)$ and $f(r,x)$. We find the family of solutions
\begin{equation}
h(r)=-2Mr+M^2-a^2+k, \>\>\>\>\>\>\> f(r,x)=b\rho^2\{\frac{(r-M)^2+kx^2}{(r-M)^2+k}\}^c
\label{3}
\end{equation}
where M, k, b and c are arbitrary constants.
The nonzero components of $R_{\mu\nu}$ of this family are the following \cite{Bo3}
\begin{equation}
R_{tt}=\frac{[(r-M)^2+a^2(1-x^2)+k](M^2-a^2+k)}{\rho^2}N
\label{5}
\end{equation}
\begin{equation}
R_{rr}=\frac{2ck\rho^2-[(r-M)^2+k](M^2-a^2+k)}{\rho^2[(r-M)^2+k]^2}
\label{6}
\end{equation}
\begin{equation}
R_{\theta\theta}=\frac{M^2-a^2+k}{\rho^2}
\label{7}
\end{equation}
\begin{equation}
R_{\phi\phi}=\frac{\{(r^2+a^2)^2+a^2[(r-M)^2+k](1-x^2)\}(1-x^2)(M^2-a^2+k)}{\rho^2}N
\label{8}
\end{equation}
\begin{equation}
R_{t\phi}=-\frac{a\{r^2+a^2+(r-M)^2+k\}(1-x^2)(M^2-a^2+k)}{\rho^2}N
\label{9}
\end{equation}
where
\begin{equation}
N=\frac{1}{b(\rho^2)^2}\{\frac{(r-M)^2+kx^2}{(r-M)^2+k}\}^{-c}
\label{10}
\end{equation}
The eigenvalues of the matrix $[(R_{t}^{t},R_{t}^{\phi}),(R_{\phi}^{t},R_{\phi}^{\phi})]$ are
\begin{equation}
\lambda_\pm=\pm(M^2-a^2+k)N
\label{11}
\end{equation}
and the other eigenvalues $\lambda_r$ and $\lambda_\theta$ of the matrix $R_{\mu}^{\nu}$ are
\begin{equation}
\lambda_r=\frac{2ck\rho^2-[(r-M)^2+k](M^2-a^2+k)}{(r-M)^2+k}N
\label{12}
\end{equation}
\begin{equation}
\lambda_\theta=\lambda_+=(M^2-a^2+k)N
\label{13}
\end{equation}
Calculating the eigenvectors of the matrix $R_{\mu}^{\nu}$ we find that the timelike eigenvector $(u_t)^\mu$ corresponds to the eigenvalue $\lambda_-$ that is
\begin{equation}
\lambda_t=\lambda_-=-(M^2-a^2+k)N
\label{14}
\end{equation}
and its normalized form is
\begin{equation}
(u_t)^\mu=\frac{1}{\sqrt{\rho^2[(r-M)^2+k]}}\{(r^2+a^2)\delta_{t}^{\mu}+a\delta_{\phi}^{\mu}\}
\label{15}
\end{equation}
(if we choose + as overall sign). Then the eigenvector $(u_\phi)^{\mu}$ corresponds to the eigenvalue $\lambda_+$, that is
\begin{equation}
\lambda_\phi=\lambda_+=(M^2-a^2+k)N
\label{16}
\end{equation}
and its normalized form is
\begin{equation}
(u_\phi)^\mu=\frac{1}{\sqrt{\rho^2(1-x^2)}}\{a(1-x^2)\delta_{t}^{\mu}+\delta_{\phi}^{\mu}\}
\label{17}
\end{equation}
Also the normalized eigenvectors $(u_r)^\mu$ and $(u_\theta)^\mu$, which correspond to the eigenvalues $\lambda_r$ and $\lambda_\theta$ respectively, are
\begin{equation}
(u_r)^\mu=\sqrt{[(r-M)^2+k]\rho^2N}\delta_{r}^{\mu}\>\>\> \mbox{and} \>\>\> (u_\theta)^\mu=\sqrt{\rho^2N}\delta_{\theta}^{\mu}
\label{18}
\end{equation}

The Ricci scalar $R$ of our solution is \cite{Bo3}
\begin{equation}
R=\frac{2ck\rho^2}{(r-M)^2+k}N=\frac{2ck[(r-M)^2+k]^{c-1}}{b\rho^2[(r-M)^2+kx^2]^c}
\label{19}
\end{equation}
The eigenvalues $w_i=\lambda_i-R/2$, $i=t,r,\theta,\phi$ of the energy-momentum tensor $T_{\mu}^{\nu}=R_{\mu}^{\nu}-\frac{R}{2}\delta_{\mu}^{\nu}$ are calculated from Eqs (\ref{12})-(\ref{14}), (\ref{16}) and (\ref{19}). We get
\begin{equation}
w_t=-\frac{(M^2-a^2+k)[(r-M)^2+k]+ck\rho^2}{(r-M)^2+k}N
\label{20}
\end{equation}
\begin{equation}
w_r=-\frac{(M^2-a^2+k)[(r-M)^2+k]-ck\rho^2}{(r-M)^2+k}N
\label{21}
\end{equation}
\begin{equation}
w_\theta=w_\phi=\frac{(M^2-a^2+k)[(r-M)^2+k]-ck\rho^2}{(r-M)^2+k}N
\label{22}
\end{equation}
The energy density $\mu$ is given by the relation
\begin{equation}
\mu=-w_t=\frac{(M^2-a^2+k)[(r-M)^2+k]+ck\rho^2}{(r-M)^2+k}N
\label{23}
\end{equation}
If we define A and B by the relation
\begin{equation}
A=(M^2-a^2+k)N \>\>\> \mbox{and} \>\>\> B=\frac{ck\rho^2}{(r-M)^2+k}N
\label{24}
\end{equation}
we get
\begin{equation}
\mu=A+B, \>\>\>\>\> w_r=-A+B, \>\>\>\>\> w_\theta=w_\phi=A-B
\label{25}
\end{equation}
It is easy to find that if
\begin{equation}
A\geq0 \>\>\>\>\> \>\> \mbox{and} \>\>\>\>\>\> B\geq0
\label{26}
\end{equation}
the above expressions for $\mu$, $w_r$, $w_\theta$ and $w_\phi$ satisfy the dominant, the strong and of course the weak  energy conditions \cite{Ha1}. Relations (\ref{26}) are satisfied, for example if
\begin{equation}
(r-M)^2+kx^2 >0\>\>\>\>\>\> \mbox{and}\>\>\>\>\>\> bck\geq0 \>\>\>\>\>\> \mbox{and} \>\>\>\>\>\> b(M^2-a^2+k)\geq0
\label{27}
\end{equation}

From Eqs (\ref{15})-(\ref{17}) we can calculate the normalized eigenvectors $(u_t)_\mu$, $(u_\phi)_\mu$, $(u_r)_\mu$ and $(u_\theta)_\mu$. We find
\begin{equation}
(u_t)_\mu=\frac{(r-M)^2+k}{\sqrt{\rho^2[(r-M)^2+k]}}\{-\delta_{t\mu}+a(1-x^2)\delta_{\phi\mu}\}
\label{28}
\end{equation}
\begin{equation}
(u_\phi)_\mu=\frac{1-x^2}{\sqrt{\rho^2(1-x^2)}}\{-a\delta_{t\mu}+(r^2+a^2)\delta_{\phi\mu}\}
\label{29}
\end{equation}
\begin{equation}
(u_r)_\mu=\frac{1}{\sqrt{[(r-M)^2+k]\rho^2N}}\delta_{r\mu}\>\>\> \mbox{and} \>\>\> (u_\theta)_\mu=\frac{1}{\sqrt{\rho^2N}}\delta_{\theta\mu}
\label{30}
\end{equation}
The energy-momentum tensor $T_{\mu\nu}$ of the solution (\ref{1})-(\ref{3}) can be calculated from Eqs (\ref{5})-(\ref{10}), (\ref{19}), (\ref{21})-(\ref{23}) and (\ref{28})-(\ref{30}). This tensor can be written in the form
\begin{equation}
T_{\mu\nu}=(\mu+w_\perp)(u_t)_\mu(u_t)_\nu +w_\perp g_{\mu\nu}+(w_\parallel-w_\perp)(u_r)_\mu(u_r)_\nu
\label{31}
\end{equation}
where
\begin{equation}
w_\perp=w_\theta=w_\phi,\>\>\>\>\>\> \mbox{and} \>\>\>\>\>\> w_\parallel=w_r
\label{32}
\end{equation}
The above $T_{\mu\nu}$ is the energy-momentum tensor of an anisotropic fluid \cite{He1}.

The infinite red shift surfaces of our family of solutions are obtained from the relation $g_{tt}=0$. From this relation we get
\begin{equation}
r_\pm^{RS}=M\pm\sqrt{a^2(1-x^2)-k}
\label{33}
\end{equation}
For $k<0$ the above surfaces are closed and axially symmetric.

The solutions have irremovable singularities at the points at which at least one of the invariants $R$ and $R^2=R_{\mu\nu\zeta\eta}R^{\mu\nu\zeta\eta}$ is singular. The Ricci scalar $R$ is given by Eq. (\ref{19}), while the curvature scalar $R^2$ was computed for certain positive and negative small values of $c$ \cite{Bo3}. For all $c$ for which $R^2$ was computed it was found that all singularities of $R^2$ are also singularities of $R$. Therefore we shall determine the singularities of the solutions from Eq. (\ref{19}). From this expression it is obvious that for any $c$ we have an irremovable singularity at
\begin{equation}
\rho^2=r^2+a^2x^2=0
\label{34}
\end{equation}
which is the well known ring singularity of Kerr's solution \cite{Ra1}, \cite{Vi1}. For $c<0$ no other singularities exist if $k>0$, while if $k<0$ the solutions are singular also if $(r-M)^2+k=0$ that is if $r=M\pm\sqrt{-k}$
If $c>0$ the solutions are in addition singular if $(r-M)^2+kx^2=0$, which for $k>0$ is satisfied for $r=M$ and $x=0$, and for $k<0$ for $r=M\pm\sqrt{-k}x$

We shall examine now if our solutions can be matched to the solution of Kerr on a closed surface $S$ with Kerr's solution as exterior solution. We express the solution of Kerr in Boyer-Lindquist coordinates \cite{Bo2}, since our family of solutions is expressed in such coordinates. We assume that the coordinates $r$ and $\theta $ of the various points of $S$  are not independent but can be expressed by a single parameter $\tau$. Therefore $S$ has the coordinates $\zeta^i=(t,\tau,\phi)$, while the space time on which it is embedded has coordinates $x^\alpha=(t,r,\theta,\phi)$. Eliminating $\tau$ we get for $S$ an equation of the form $r=R(\theta)$.
It is easy to show that the 3-metric $^3g_{ij}$ of the surface is connected with the 4-metric $^4g_{\alpha\beta}$ of the spacetime by the relation
\begin{equation}
^3g_{ij}=\frac{\partial x^\alpha}{\partial \zeta^i}\frac{\partial x^\beta}{\partial \zeta^j}{^4g_{\alpha\beta}}
\label{36}
\end{equation}
where Latin indices take the values $(t,\tau,\phi)$ and Greek indices the values $(t,r,\theta,\phi)$. For any metric dependent quantity $P$ we must specify the region in which it is calculated. The notation $P^{+}$($P^{-}$) means that P is calculated in the exterior (interior) region of S. The notation  $P^{+}|_S$ ($P^{-}|_S$) means that the quantity $P$ is calculated in the exterior (interior) region and evaluated at the surface, while we use the notation
\begin{equation}
[P]\equiv P^{+}|_S-P^{-}|_S
\label{37}
\end{equation}
which means that [P] denotes the discontinuity of $P$ at the surface.

The Darmois-Israel conditions \cite{Da1},\cite{Is1} for the matching of the interior and exterior regions are continuity of the first fundamental form
\begin{equation}
[^3g_{ij}]=0
\label{38}
\end{equation}
and continuity of the extrinsic curvature $K_{ij}$ ( second fundamental form)
\begin{equation}
[K_{ij}]=0
\label{39}
\end{equation}
If both conditions are satisfied we refer to $S$ as boundary surface. If only condition (\ref{38}) is satisfied we refer to $S$ as thin shell. If the only off diagonal term of the metrics to be joined is $g_{t\phi}$
condition (\ref{38}) implies the relations \cite{Dr1}
\begin{equation}
[^3g_{tt}]=[^4g_{tt}]=0, \>\>\>\> [^3g_{t\phi}]=[^4g_{t\phi}]=0, \>\>\>\> [^3g_{\phi\phi}]=[^4g_{\phi\phi}]=0
\label{40}
\end{equation}
\begin{equation}
[^3g_{\tau\tau}]=(\frac{\partial r}{\partial \tau})^2[^4g_{rr}]+(\frac{\partial \theta}{\partial \tau})^2[^4g_{\theta\theta}]=0
\label{41}
\end{equation}
Eqs (\ref{40}) are satisfied if
\begin{equation}
k=a^2-M^2
\label{42}
\end{equation}
while   Eq. (\ref{41}) is satisfied if as matching surfaces  we chose the surfaces
\begin{equation}
b\{\frac{(r^S-M)^2+kx^2}{(r^S-M)^2+k}\}^c-1=0
\label{43}
\end{equation}
or the surface
\begin{equation}
r^S-M=\sqrt{M^2-a^2}\cos(\theta-\theta_0)
\label{44}
\end{equation}
where $\theta_0$ is an arbitrary constant.Assuming that
\begin{equation}
y\equiv\frac{a^2}{M^2}<1 \>\>\>\>\>\> \mbox{and} \>\>\>\>\>\> v\equiv\sqrt[c]{b}<1
\label{44'}
\end{equation}
we get from Eq. (\ref{43}) the matching surfaces $S_+$ and $S_-$
\begin{equation}
r_\pm^S-M=\pm M\sqrt{\frac{1-y}{1-v}}\sqrt{1-vx^2}
\label{45}
\end{equation}
where the surface  $S_+$ ( $S_-$ ) is given by the above expression with the plus (minus) sign. In Boyer-Lindquist coordinates \cite{Bo2} an equation of the form $r=constant$ is an oblate spheroid \cite{Gu1}. Therefore if $v\ll y$ our matching surfaces $S_+$ and $S_-$ approximate oblate spheroids. Also the matching surface of Eq. (\ref{44}) approximetes an oblate spheroid if $1-y\ll 1$.

Condition (\ref{39}) implies the relations \cite{Dr1}. \cite{Ky3}
\begin{equation}
[K_{tt}]=[g^{rr}]g_{tt,r}-[g^{\theta\theta}]R_{,\theta}g_{tt,\theta}=0
\label{46}
\end{equation}
\begin{equation}
[K_{t\phi}]=[g^{rr}]g_{t\phi,r}-[g^{\theta\theta}]R_{,\theta}g_{t\phi,\theta}=0
\label{47}
\end{equation}
\begin{equation}
[K_{\phi\phi}]=[g^{rr}]g_{\phi\phi,r}-[g^{\theta\theta}]R_{,\theta}g_{\phi\phi,\theta}=0
\label{48}
\end{equation}
\[[K_{\tau\tau}]=\frac{1}{2}(\frac{\partial r}{\partial \tau})^2\{[g^{rr}g_{rr,r}]+R_{,\theta}[g^{\theta\theta}g_{rr,\theta}]\}+\frac{\partial r}{\partial \tau}\frac{\partial \theta}{\partial \tau}\{[g^{rr}g_{rr,\theta}]-R_{,\theta}[g^{\theta\theta}g_{\theta\theta,r}]\}\]
\begin{equation}
-\frac{1}{2}(\frac{\partial \theta}{\partial \tau})^2\{[g^{rr}g_{\theta\theta,r}]+R_{,\theta}[g^{\theta\theta}g_{\theta\theta,\theta}]\}=0
\label{49}
\end{equation}
where all metric components refer to the 4-metric and we have used the notation $P_{,a}=\frac{\partial P}{\partial x^a}$. Eqs $[K_{t\tau}]=[K_{\phi\tau}]=0$ are identically satisfied.
For $k$ and $S_\pm$ given by Eqs (\ref{42}) and (\ref{45}) respectively we find that
\begin{equation}
[K_{tt}]=[K_{t\phi}]=[K_{\phi\phi}]=0
\label{50}
\end{equation}
Also taking $\tau=\theta$ we find for the surfaces $ S_+ $ and $ S_- $ of  Eqs (\ref{45})
\begin{equation}
[K_{\tau\tau}]=[K_{\theta\theta}]=\mp\frac{cM\sqrt{(1-y)(1-v)}}{\sqrt{(1-vx^2)^3}}
\label{51}
\end{equation}
where the minus (plus) sign corresponds to $S_+$ ($S_-$). Therefore both surfaces $S_+$ and $S_-$ are thin shells. The surface of Eq (\ref{44}) is again thin shell \cite{Ky4}.

 To calculate the surface energy tensor $S_i^j$ we use the Lanczos relation \cite{Is2}, \cite{Is3},  \cite{La1}
\begin{equation}
-8\pi S_i ^j=[K_{il}]^3g^{lj}-\delta_i^j([K_{ln}]^3g^{ln})
\label{52}
\end{equation}
For $^3g_{ij}$ and $K_{ij}$ given by Eqs (\ref{36}), (\ref{50}) and (\ref{51}) we find that the only non-vanishing $S_i^j$ are the following
\begin{equation}
S_t ^t=S_\phi^\phi=\frac{1}{8\pi} [K_{\theta\theta}]^3g^{\theta\theta}= \frac{1}{8\pi}\frac{1}{^4g_{\theta\theta}+^4g_{rr}(\frac{\partial r}{\partial \theta})^2 }[K_{\theta\theta}]
\label{53}
\end{equation}
The  surface density $\sigma(\theta)$ is defined by the eigenvalue equation
\begin{equation}
 S_b ^au^b=-\sigma u^a \>\>\>\>\> \mbox{with} \>\>\>\>\>\> u_au^a=-1
\label{54}
\end{equation}
From Eqs (\ref{51}), (\ref{53}) and (\ref{54}) we get
\[  \sigma(\theta)=-\frac{1}{8\pi}\frac{1}{^4g_{\theta\theta}+^4g_{rr}(\frac{\partial r}{\partial \theta})^2 }[K_{\theta\theta}] =\pm\frac{cM}{8\pi\rho^2}\sqrt{\frac{(1-y)(1-v)}{1-vx^2}}\]
\begin{equation}
=\pm\frac{c}{8\pi M}\{(1+\{\frac{(1-y)(1-vx^2)}{1-v}\}^\frac{1}{2})^2+yx^2\}^{-1}\sqrt{\frac{(1-y)(1-v)}{1-vx^2}}
\label{55}
\end{equation}
where the signs plus and minus correspond to the matching surfaces $S_+$ and $S_-$ of Eqs (\ref{45}) respectively.

 We want to have $\sigma(\theta)>0$ and also in the interior region and on the matching surface the internal solution to satisfy the energy conditions. These conditions are satisfied if Eqs (\ref{27}) hold. Therefore since we want to have $k=a^2-M^2<0$ and $b>0$ the second of Eqs (\ref{27}) gives $c<0$. Then from Eq (\ref{55}) we conclude that in order to have $\sigma(\theta)>0$ we must take $S_-$ as matching surface. We shall make this choice. Inside this surface $r$ changes from $0$ to $M-M\sqrt{\frac{1-y}{1-v}}\sqrt{1-vx^2}$. In this range of $r$ the expression $(r-M)^2+kx^2$ for fixed $x^2$ decreases monotonically with $r$ taking values from $M^2(1-x^2) + a^2x^2 \geq 0$, which corresponds to $r=0$, to $\frac{(M^2-a^2)(1-x^2)}{1-v}\geq0$, which corresponds to $r=M-M\sqrt{\frac{1-y}{1-v}}\sqrt{1-vx^2}$. Therefore with $S_-$ as matching surface all energy conditions are satisfied in the interior region of $S_-$ and on $S_-$.

The outer event horizon $r^H_+$  of the exterior solution of Kerr is given by the relation
\begin{equation}
 r^H_\pm=M + M\sqrt{1-y}
\label{56}
\end{equation}
From Eqs (\ref{45}) and (\ref{56}) we get
\begin{equation}
 r^S_-<r^H_+
\label{57}
\end{equation}
Therefore our matching surface $S_-$ is inside the outer event horizon and of course not in the ergosphere, which is outside the outer event horizon..

To find if $S_-$ is timelike, null or spacelike we shall calculate a vector $\eta_\alpha$, which is perpendicular to it. From  Eq (\ref{45}) we find that \begin{equation}
\eta_\alpha=\{0,\>\>\> 1,\>\>\>  \frac{M v x \sqrt{(1-y)(1-x^2)}}{\sqrt{(1-v)(1-vx^2)}}, \>\>\>0\}
\label{58}
\end{equation}
from which we get
\begin{equation}
\eta_\alpha\eta^\alpha=\frac{M^2v(1-y)(1-x^2)}{((r^S_-)^2+a^2x^2)(1-v)(1-vx^2)}
\label{59}
\end{equation}
Therefore for $x^2 \neq 1$ we get $\eta_\alpha\eta^\alpha > 0$, which means that the matching surface $S_-$ is a spacelike surface.


\begin{thebibliography}{99}
\bibitem{Ky2} Details of these long calculations will be presented elsewhere
\bibitem{Ky3} Since the right hand side of Eqs (\ref{46})-(\ref{49}) is zero a positive normalization factor of the unit vector, with the covariat derivative of which we define the extrinsic curvature, is ommitted.  We shall ommit this factor everywhere in the paper.
\end{thebibliography}
\end{document}